\documentclass[12pt]{article}
\usepackage{epsf,amsfonts,amssymb,epsfig,amsmath,mathtools,graphics,slashed}
\usepackage{hyperref,hep,graphicx,subfig}
\usepackage{rotating}
\usepackage{xcolor}
\usepackage{verbatim}
\usepackage[final]{showlabels}

\definecolor{greenish}{RGB}{0,190,0}

\definecolor{yellowish}{RGB}{190,190,0}

\definecolor{bluish}{RGB}{0,0,190}

\newcommand{\nn}{\notag \\}

\makeatletter

\@addtoreset{equation}{section}
\makeatother

\begin{document}

\begin{titlepage}

\vfill


\vfill

\begin{center}
   \baselineskip=16pt
   {\Large\bf Nearly Critical Superfluids in Keldysh-Schwinger Formalism}
  \vskip 1.5cm
  \vskip 1.5cm
      Aristomenis Donos and Polydoros Kailidis\\
   \vskip .6cm
   \begin{small}
      \textit{Centre for Particle Theory and Department of Mathematical Sciences,\\ Durham University, Durham, DH1 3LE, U.K.}
   \end{small}\\            
\end{center}

\vfill

\begin{center}
\textbf{Abstract}
\end{center}
\begin{quote}
We examine the effective theory of critical dynamics near superfluid phase transitions in the framework of the Keldysh-Schwinger formalism. We focus on the sector capturing the dynamics of the complex order parameter and the conserved current corresponding to the broken global symmetry. After constructing the theory up to quadratic order in the $a$-fields, we compare the resulting stochastic system with Model F as well as with holography. We highlight the role of a time independent gauge symmetry of the effective theory also known as ``chemical shift". Finally, we consider the limiting behaviour at energies much lower than the gap of the amplitude mode by integrating out the high energy degrees of freedom to reproduce the effective theory of superfluids.
\end{quote}

\vfill

\end{titlepage}

\setcounter{equation}{0}
\section{Introduction}
The dynamics of nearly critical systems is a fascinating topic with a long history. Early theories of critical phenomena aiming to understand the small relaxation rates in systems close to criticality, started with the conventional theory of critical slow down \cite{PhysRev.95.1374,1965626}. This turned out to have limited range of applicability \cite{KAWASAKI19671277} to real systems and a significant improvement was the mode-coupling theories \cite{doi:10.1063/1.1732501,PhysRev.150.291,KAWASAKI19701,PhysRev.166.89}. Mode-coupling took into account the coupling of the order parameter to other hydrodynamic slow modes that exist at wavelengths larger than the correlation length of the system. The dynamics of the coupled system is described by stochastic equations with Gaussian noise. In parallel to static renormalisation group techniques for systems undergoing a phase transition \cite{1974PhR}, a systematic treatment of interactions was carried out in \cite{RevModPhys.49.435}.

 Very recently, a systematic study of low energy effective field theories was proposed in the context of the Keldysh-Schwinger closed time path formalism \cite{Haehl:2015uoc,Crossley:2015evo}. The fluctuation dissipation theorem is built in the formalism due to the Kubo-Martin-Schwinger symmetry \cite{Kubo:1957mj,Martin:1959jp} providing an appropriate framework to consider thermal fluctuations beyond Gaussian noise \cite{Chen-Lin:2018kfl,Jain:2020zhu,Jain:2020hcu}.  This is in parallel to the standard approach of stochastic systems \cite{10.1143/PTP.33.423} where the variance of the noise fields is fixed again in a way that the probability distribution of the system will always relax to the Boltzmann distribution \cite{RevModPhys.32.25} and the fluctuation dissipation theorem will be satisfied.

In this paper we will construct an effective action for systems close to a superfluid phase transition up to quadratic order in the $a$-fields. As we will see, at this order in the $a$-fields, our system is essentially described by Model F in the classification of Hohenberg and Halperin \cite{RevModPhys.49.435}. Setting all our external $a$-sources to zero will be necessary as the equations for stochastic hydrodynamics are written in terms of $r$-field sources.

An interesting aspect of our construction is that by choosing an appropriate set of dynamical variables, our effective theory can be viewed as a simple coupling between the $U(1)$ symmetric version of Model A and the diffusive model for current conservation \cite{Crossley:2015evo,Liu:2018kfw}. Both models possess their own global $U(1)$ symmetry which in the case of charge diffusion is gauged by the external electromagnetic field. Interestingly, the diffusive model for current conservation possesses an additional time independent gauge symmetry, also known as the ``chemical shift'' symmetry \cite{Dubovsky:2011sj,Crossley:2015evo}. Our final effective theory can be obtained by only imposing the diagonal part of the two global $U(1)$ symmetries and gauging the symmetry of Model A by the time independent gauge symmetry of charge diffusion.

By considering fluctuations of frequencies much lower than the gap of the amplitude mode, we end up with an effective description of the conserved current in the broken phase. At the level of classical hydrodynamics, this system is parametrised by the charge and current susceptibilities as well as the third bulk viscosity and the incoherent conductivity \cite{KHALATNIKOV198270,Herzog:2011ec,Bhattacharya:2011tra,Damle:1997rxu}. As one might expect, the low energy theory is described by the phase of the complex order parameter. We express the third bulk viscosity of the superfluid in terms of bare thermodynamic quantities and the complex kinetic coefficient of the order parameter.

\section{The Effective Theory}\label{sec:formalism}

The basic ingredient of the Keldysh-Schwinger formalism \cite{Schwinger:1960qe,Keldysh:1964ud} is the acceptance that time evolution of quantum fields has to happen along a closed time path \cite{Schwinger:1960qe}. In its full generality, the formalism can be used to extract powerful statements about the dynamics of quantum systems around a state captured by a generic density matrix. More recently, significant progress has been made in realising  effective theories describing the long wavelength, small frequency limit within the same framework \cite{Crossley:2015evo}. After integrating out the fast, short wavelength modes of the system in the closed time path integral we are left with the effective theory fields $\chi$ and the sources $\phi$. In terms of these, the closed time path generating functional reads,
\begin{align}\label{eq:generating_eft}
e^{W[\phi_1,\phi_2]}=\int D\chi_1\,D\chi_2\,e^{i\,I_{EFT}[\chi_1,\phi_1;\chi_2,\phi_2]}\,,
\end{align}
where $I_{EFT}$ is the effective action, with $\chi_1$ and $\chi_2$ corresponding to the forward and backward time evolution branch respectively. In order for the path integral in \eqref{eq:generating_eft} to represent the trace of a closed time path, the dynamical fields $\chi_1$ and $\chi_2$ must satisfy the gluing conditions,
\begin{align}\label{eq:bc_infinity}
\lim_{t\rightarrow +\infty}\left(\chi_1(t)-\chi_2(t)\right)=\lim_{t\rightarrow -\infty}\left(\chi_1(t)-\chi_2(t-i\beta)\right)=0\,,
\end{align}
where $\beta$ is the inverse temperature of the thermal state. Due to the integration over the UV modes, the action $I_{EFT}$ will carry dependence on the thermodynamic variables fixing the thermal state. 

It is convenient to introduce the $r$ and $a$-fields through,
\begin{align*}
    \chi_r=\frac{1}{2}\left(\chi_1+\chi_2 \right)\,,\quad \chi_a=\chi_1-\chi_2\,,
\end{align*}
and similarly for the sources $\phi_r$ and $\phi_a$. The constraints implied by unitarity read \cite{Crossley:2015evo},
\begin{align}\label{eq:I_eft_properties_ar}
    &I_{EFT}^\star[\chi_r,\phi_r;\chi_a,\phi_a]+I_{EFT}[\chi_r,\phi_r;-\chi_a,-\phi_a]=0\,,\notag\\
    &I_{EFT}[\chi_r,\phi_r;\chi_a=0,\phi_a=0]=0\,,\notag\\
    &\mathrm{Im}\,I_{EFT}[\chi_r,\phi_r;\chi_a,\phi_a]\geq 0\,.
\end{align}

When both the Hamiltonian of our system and the thermal state are invariant under a symmetry group $G$, we would anticipate that the effective theory will possess the same symmetry. Due to the doubling of fields, one might suspect that the symmetry group should be enhanced to $G\times G$. This is true for time dependent gauge transformations which preserve the boundary conditions \eqref{eq:bc_infinity}. However, for time independent symmetry transformations, it is only the diagonal subgroup that preserves these boundary conditions. This key difference will play a crucial role in our construction.

The first degree of freedom that we need to include in our effective theory is the complex order parameter $\psi_i$ which is charged under a global $U(1)$ transformation. Moreover, we assume that this global symmetry is gauged by the external electromagnetic potential $A_{i\,\mu}$. The second observable we need to include is the associated Noether current $J_\mu$ due to the $U(1)$ symmetry and which couples to the order parameter $\psi$. In the context of the effective theory it contributes a scalar degree of freedom $\phi_i$, the Stueckelberg field of the electromagnetic gauge transformations \cite{Crossley:2015evo}.

In the normal phase, the scalar $\phi_i$ describes the diffusive dynamics of charge density \cite{Crossley:2015evo}. For this reason, the effective field theory must also be invariant under the separate diagonal time independent gauge transformation \footnote{A recent construction \cite{Kapustin:2022iih} of critical models  for superfluidity has used a similar condition. However, the reasoning in footnote 4 of \cite{Kapustin:2022iih} would, in principle, allow for the extra terms we discuss below equation \eqref{eq:fe_expansion}. Our model agrees with \cite{Kapustin:2022iih} as long as we impose the constraint $\kappa_0=-q\,c_6^I$.} \cite{Dubovsky:2011sj,Crossley:2015evo},
\begin{align}\label{eq:gtrans_t_indep}
\phi_r^\prime=\phi_r+\sigma\,,\quad\phi_a^\prime=\phi_a\,,\quad \partial_t\sigma=0\,.
\end{align}
Note that this is a distinct transformation from the electromagnetic gauge transformations as it does not involve neither the order parameter nor the external gauge field. This is known as a ``chemical shift'' symmetry \cite{Dubovsky:2011sj} and it is imposed in normal phase effective theories in order to exclude the presence of supercurrents. However, in nearly critical systems this symmetry is stronger than imposing the absence of supercurrents in their normal phase. Moreover, it will be necessary in order to match with both Model F \cite{RevModPhys.49.435} as well as with the equivalent holographic systems \cite{Donos:2022qao}.

It will be useful to write down the symmetry transformations that our effective theory should be invariant under in the notation of the $r$- and $a$-fields. Keeping only the leading order terms in the $a$-fields we obtain,
\begin{align}\label{eq:trans_col}
\psi'_r&=e^{iq \lambda_D}\psi_r\,, \quad \psi'_a=e^{iq \lambda_D}(\psi_a+iq\lambda_A \psi_r)\,,\nn
\phi'_r&=\phi_r+\lambda_D+\sigma\,, \quad \phi'_a=\phi_a+\lambda_A\,,\nn
A'_{r \mu}&=A_{r \mu}-\partial_\mu \lambda_D\,, \quad  A'_{a \mu}=A_{a \mu}-\partial_\mu\lambda_A,
\end{align}
where we have defined the diagonal $\lambda_D=\frac{1}{2}(\lambda_1+\lambda_2)$ and the anti-diagonal $\lambda_A=\lambda_1-\lambda_2$ gauge transformation parameters. The transformations of \eqref{eq:trans_col} have been linearised in $\lambda_A$ since the latter is of order $\mathcal{O}(\beta)$ in the semiclassical limit we are interested in. Given our discussion around symmetries, the gauged symmetry should exist on the forward as well as on the backward branch of the time evolution path. However, we would only have to worry about the global part of the diagonal $\lambda_D$. Finally, we will also consider source fields $s_{r,a}$ for the charged scalar transforming in the same way with the fields $\psi_{r,a}$ in \eqref{eq:trans_col}.

The appearance of more that one field transforming non-trivially under the gauge transformations allow us to write down the gauge invariant combinations,
\begin{align}\label{eq:gauge_invariant_vars}
 \hat{\psi}_r&=e^{-i q \phi_r}\psi_r\,, \qquad \hat{\psi}_a=e^{-iq\phi_r}(\psi_a-iq\phi_a \psi_r)\,,\nn
B_{r \mu}&=\partial_\mu \phi_r+A_{r \mu}\,, \qquad B_{a \mu}=\partial_\mu \phi_a+A_{a \mu}\,.
\end{align}
A natural choice for the local chemical potential is the time component of the gauge invariant $r$-vector component $\mu=B_{r\,0}$ which coincides with the thermodynamic chemical potential in equilibrium. For spacetime dependent configurations, this choice will fix our hydrodynamic frame. In terms of our gauge invariant variables, the time independent gauge transformation \eqref{eq:gtrans_t_indep} takes the form,
\begin{align}\label{eq:gtrans_t_indep_newvars}
\hat{\psi}'_r&=e^{-i q \sigma}\hat{\psi}_r\,, \quad \hat{\psi}'_a=e^{-i q \sigma}\hat{\psi}_a\,,\nn
B'_{r i}&=B_{r i}+\partial_i \sigma\,,\quad B'_{r t}=B_{r t}\,,\quad B_{a \mu}^\prime=B_{a \mu}\,.
\end{align}
leading to the natural covariant derivative,
\begin{align}\label{eq:cov_dervs}
    D_i\hat{\psi}_{r,a}&=\partial_i\hat{\psi}_{r,a}+i\,q B_{r\,i}\,\hat{\psi}_{r,a}\,,\quad D_t\hat{\psi}_{r,a}=\partial_t\hat{\psi}_{r,a}\,.
\end{align}

The above discussion shows that the effective theory we are after is essentially a combination of the $U(1)$ symmetric Model A with the charge diffusion model of \cite{Crossley:2015evo}. The degrees of freedom of the two systems are then coupled because the diagonal parts of the respective global $U(1)$ symmetries are identified. Moreover, the global $U(1)$ symmetry of the Model A sector is gauged by the time independent gauge parameter of charge diffusion.

In addition to the continuous symmetry transformations, our effective theory should also be invariant under a set of KMS transformations. Our thermal state involves a finite chemical potential, making simpler to consider the discrete transformation $\Theta=\hat{P}\,\hat{T}$ involving parity and time reversal. The KMS transformations then read\footnote{For notational simplicity, we only give the parity transformation rules in odd number of spatial dimensions. In general, we should only be flipping the sign of only one spatial coordinate. Our conclusions would remain unchanged.},
\begin{align}\label{eq:kms_higgs}
\tilde{B}_{r \mu}(-x)&=B_{r\mu}(x)\,,\quad \tilde{B}_{a \mu}(-x)=B_{a\mu}(x)+i\beta\,\partial_t B_{r\mu}(x)\,,\nn
\tilde{\hat{\psi}}_r(-x)&=\hat{\psi}^\star_r(x)\,,\quad
\tilde{\hat{\psi}}_a(-x)=\hat{\psi}^\star_a(x) +i\beta\, \partial_t \hat{\psi}^\star_r (x)\,.
\end{align}

The KMS transformation rules \eqref{eq:kms_higgs} preserve the total number of $a$-field factors and time derivatives of $r$-fields. It therefore makes sense to write an expansion for the effective Lagrangian density according to,
\begin{align}\label{eq:L_EFT_exp}
    \mathcal{L}_{EFT}=\mathcal{L}^{[1]}+\mathcal{L}^{[2]}+\cdots\,,
\end{align}
where the term $\mathcal{L}^{[n]}$ contains terms with a total of $n$ factors of $a$-fields and time derivatives of $r$-fields. In this notation, the second line of equation \eqref{eq:I_eft_properties_ar} guarantees that $\mathcal{L}^{[0]}$ vanishes.

Here, we will give the highlights of a more detailed argument for the construction, which can be found in Appendix \ref{app:eff_action_con}. Keeping only linear terms in the scalar field sources which are invariant under the symmetry as well as the KMS transformations, the leading term in \eqref{eq:L_EFT_exp} takes the form,
\begin{align}\label{eq:L_1_exp}
\mathcal{L}^{[1]}=&2\,\mathrm{Re}\left[e^{iq\phi_r}s_r^\star\,\hat{\psi}_a+e^{-iq\phi_r}\,\hat{\psi}_r^\star s_a+iq\,e^{iq\phi_r}\,\hat{\psi}_r\,s_r^\star\,\phi_a\right]\nn
&+\frac{\delta S_0}{\delta B_{r\,\mu}}\,B_{a\,\mu}+2\,\mathrm{Re}\left( \frac{\delta S_0}{\delta \hat{\psi}_r}\,\hat{\psi}_a\right)\,.
\end{align}
The functional $S_0$ takes the general form,
\begin{align}
    S_0[\hat{\psi}_r,\hat{\psi}_r^\star,\phi_r;A_{r\,\mu}]=- \int d^{d-1} x\, F\,,
\end{align}
with $F$ being a function of quantities which are invariant under the transformation \eqref{eq:gtrans_t_indep_newvars}, such as $\hat{\psi}_r\,\hat{\psi}^\star_r$, $D_i\hat{\psi}_r \,D^i\hat{\psi}^\ast_r$ and $B_{r\,t}$.

Close to the phase transition, the function $F$ should be close to the free energy density $F_0$ of the normal phase. Introducing a perturbative parameter $\varepsilon$
and assuming the scaling behaviour\footnote{ $\varepsilon$ parametrises the distance from the critical point. In this paper we want to focus on the small wavevector, small frequency response of the nearly critical system, and this is why we take the spacetime derivatives to be of the same order. The time derivative can only be $\mathcal{O}(\varepsilon^2)$, as can be seen from matching the $\varepsilon$ powers across different terms of the action.}

\begin{align}\label{eq:epsilon_exp}
    \partial_t\propto\mathcal{O}(\varepsilon^2)\,,\quad \partial_i\propto |\hat{\psi_r}|\propto\mathcal{O}(\varepsilon)\,,\quad
    \phi_r\propto \mathcal{O}(\varepsilon^0)\,.
\end{align}
we can write an expansion,
\begin{align}\label{eq:fe_expansion}
    F&=F_0-\rho_n\,B_{r\,t}-\frac{1}{2}\chi_n\,B_{r\,t}^2+r_0\,|\hat{\psi}_r|^2+\frac{1}{2}u_0\,|\hat{\psi}_r|^4\nn
    &+\kappa_0\,B_{r\,t}\,|\hat{\psi}_r|^2+w_0\,D_i\hat{\psi}_r\,D^i\hat{\psi}^\star_r\,,
\end{align}
with $u_0$, $w_0$ and $\kappa_0$ finite while $r_0\propto \mathcal{O}(\varepsilon^2)$. Moreover, we have shifted $B_{r\,t}$ by the chemical potential of the thermal state $\mu_0$ and all the bare constants appearing in \eqref{eq:fe_expansion} are now functions of $\beta$ and $\mu_0$. The symmetry \eqref{eq:gtrans_t_indep} is stronger than demanding that the effective theory should be independent of purely spatial derivatives of $\phi_r$ in the absence of the condensate. This would allow for terms of the form $|\psi_r|^2\,B_{r\,i}\,B_r{}^i$ and $B_{r}{}^i\,\mathrm{Im}(\hat{\psi}^\star_r\,D_i\hat{\psi}_r)$ which are again of order $\mathcal{O}(\varepsilon^4)$ and vanish in the normal phase.

For the second term in the expansion \eqref{eq:L_EFT_exp} we find the gauge invariant expression,
\begin{align}\label{eq:L_2_exp}
\mathcal{L}^{[2]}=&-2\,i\,\beta^{-1} c_6^R \,\hat{\psi}_a\hat{\psi}_a^\star+c_6\, \hat{\psi}_a\, \partial_t \hat{\psi}_r^\star+c_6^\star\, \hat{\psi}_a^\star\, \partial_t\hat{\psi}_r\notag\\
&-i \beta^{-1} c_5\, B_{a\, i}^2+c_5\, B_{a\, i}\, \partial_t B_{r\, i}\,,
\end{align}
which also respects the KMS symmetry \eqref{eq:kms_higgs}. Moreover, the third condition of constraints \eqref{eq:I_eft_properties_ar} implies that $c_6=c_6^R+i\, c_6^I$ can be complex with $c_6^R\leq0$ and $c_5\leq0$. Our effective action is invariant under charge conjugation $\hat{C}$ since as we show in Appendix \ref{app:eff_action_con}, e.g. the constant $\kappa_0$ is odd under $\hat{C}$.

Finally, the Noether current of the global symmetry that rotates the $a$ fields can be obtained be differentiating the effective action with respect to the $a$-vector fields,
\begin{align}\label{eq:currents_offshell}
   \hat{J}_r^i&=-2q\,\mathrm{Im}\left(\hat{\psi}^\star_r\,\frac{\partial F}{\partial D_i\hat{\psi}^\star_r}\right)+c_5\left(\partial^i\mu-E^i\right)-i\frac{2\,c_5}{\beta}B_{a}{}^i\,,\notag\\
    \hat{J}_r^t&=-\frac{\partial F}{\partial B_{r\,t}}\,,
\end{align}
where $E_i=\partial_i A_{r\,t}-\partial_t A_{r\,i}$ is the external electric field.

\section{Model F from Keldysh-Schwinger}\label{sec:stochastic}
An interesting fact about Keldysh-Schwinger effective theories at quadratic level in $a$-fields is that they are equivalent to a set of kinetic equations for the $r$-fields \cite{Crossley:2015evo}. In particular, we will recast our theory as the stochastic Model F of \cite{RevModPhys.49.435} which describes the nearly critical dynamics of a complex order parameter and the corresponding conserved current.

As we show in Appendix \ref{app:stochastic}, by using standard path integral techniques, we can recast the path integral \eqref{eq:generating_eft} in terms of noise fields instead of the $a$-fields. More specifically, we can trade $\hat{\psi}_a$ and $B_{a\,\mu}$ for the complex noise field $z$ and the real vector noise field $\zeta_\mu$ respectively following a Gaussian distribution with zero mean. 

Written in terms of the function $F$ and the chemical potential $\mu$, the system of stochastic equations becomes,
\begin{align}\label{eq:stochastic_sys_v2}
     c_6^\star\, \partial_t \hat{\psi}_r&=\frac{\partial F}{\partial \hat{\psi}_r^\star}-D_i\left(\frac{\partial F}{\partial D_i\hat{\psi}^\star_r}\right)-e^{-iq\phi_r}\,s_r-z\,,\nn
    \partial_\mu J^\mu_r&=\mathrm{Im}\left[ \hat{\psi}^\star_r\,e^{-iq\phi_r}\,s_r\right]-\partial_i\zeta^i\,,
\end{align}
where the current is given by \eqref{eq:currents_offshell} with trivial $B_{a\,i}$. The real and imaginary parts of $z$ and the components $\zeta_i$ have variance $-c_6^R/\beta$ and $-2\,c_5/\beta$ respectively. In the equations above, we have set the $a$-sources  equal to zero since we are interested in comparing with a theory which is meant to compute only the retarded Green's functions and the $r$-sources are sufficient for this purpose.

We now turn our attention to Model F of \cite{RevModPhys.49.435} which we will match with the stochastic system \eqref{eq:stochastic_sys_v2}. The effective degrees of freedom in that description are the charge density excitation $m$ above the normal phase and the complex order parameter $\psi$ which satisfy the stochastic system of equations,
\begin{align}\label{eq:HH_eoms}
    \partial_t\psi&=-2\,\Gamma_0\,\frac{\delta W}{\delta\psi^\star}-i\,g_0\,\frac{\delta W}{\delta m}\,\psi+\theta\,,\notag\\
    \partial_t m&=\lambda_0^m\,\partial_i\partial^i\left(\frac{\delta W}{\delta m} \right)+2g_0\,\mathrm{Im}\left[\psi^\star\,\frac{\delta W}{\delta\psi^\star}\right]+\zeta_H\,,
\end{align}
where $\Gamma_0$ is a complex parameter with positive real part, $\lambda_0^m$ is a positive constant, $\theta$ and $\zeta_H$ are noise fields. More specifically, the real and imaginary parts of the random field $\theta$ and the field $\zeta_H$ follow Gaussian distributions of zero mean and variances $\mathrm{Re}\,\Gamma_0/\beta$ and $-2\lambda_0^m/\beta\,\partial_i\partial^i$ respectively \cite{Halperin:1974zz}.

The variance of the $z$-distribution was fixed by the KMS conditions. In contrast, the variance of the field $\theta$ for Model F, is fixed by demanding that the fluctuation-dissipation theorem is satisfied. Equivalently, the relevant Fokker-Planck equation admits the grand canonical probability distribution $P[\psi]=Z^{-1}\,e^{-\beta\,W[\psi]}$ as a fixed point. The latter highlights that $W$ plays the role of Ginzburg-Landau-Wilson free energy.

Assuming that the free energy $W$ depends analytically on the fields $\psi$, $m$ and the corresponding classical sources $h$ and $h_m$, we can write the leading order terms,
\begin{align}
    W=&W_0-\int d^{d-1}x\,\left[ h_m\,m+\mathrm{Re}(h\,\psi^\star)\right]\,,\notag\\
    W_0=&\int d^{d-1}x\,\left[\frac{1}{2}\tilde{r}_0\,|\psi|^2 +\frac{1}{2}\tilde{w}_0\, |\nabla\psi|^2+\tilde{u}_0\,|\psi|^4\right.\nn&+\left.\frac{1}{2\,C_0}\,m^2+\gamma_0\,m\,|\psi|^2\right]\,.
\end{align}
We note that \cite{RevModPhys.49.435} has $\tilde{w}_0=1$ but this is not a significant difference as $w_0$ can be set to unity via a field redefinition.

In order to compare the stochastic systems \eqref{eq:stochastic_sys_v2} and \eqref{eq:HH_eoms}, we need to perform a change of variables from the charge density difference $m$ to the chemical potential $\mu_{h}=\frac{\delta W_0}{\delta m}$. To achieve this, we consider the Legendre transformation of the energy potential $W_0$ according to $\tilde{W}_0=W_0-\int \mu_{h}\,m$.

To match the system of equations \eqref{eq:HH_eoms} to the one obtained from the Keldysh-Schwinger effective action \eqref{eq:stochastic_sys_v2}, we identify,
\begin{align*}
    \psi&=e^{iq\phi_r}\hat{\psi}_r\,,\quad\mu_h=\mu\,,\quad h=2\,s_r\,,\quad  h_m=A_t\,,\nn
    \,A_i&=0\,,\quad \,z=-c_6^\star\,e^{-iq\phi_r}\,\theta\,,\quad \zeta_H=-\partial_i\zeta^i\,,\quad q=-g_0\,.
\end{align*}
Then the matching requires that,
\begin{align}\label{eq:model_f}
\tilde{W}_0=&\int d^{d-1}x\,\left(F-F_0(\mu_0)+\rho_n\,B_{r\,t}\right)\,,
\nn c_6^\star=&-\frac{1}{2\,\Gamma_0}\,,\quad c_5=-\lambda_0^m\,.
\end{align}

In our recent paper \cite{Donos:2022qao}, we matched the dynamics of Model F with a suitable class of holographic models of superfluidity.  At the level of mean field theory, we showed that the holographic theories capture the same dynamics near criticality with Model F. However, given that holography provides a weakly coupled description of strongly coupled microscopic theory, the transport coefficients $\Gamma_0$ and $\lambda_0^m$ (or equivalently $c_6$ and $c_5$) where fixed in terms of black hole horizon invariants.

\section{Superfluid Hydrodynamics at Low Energies}\label{sec:superfluidity}

In this section we will consider the limit of our effective theory at energies much lower than the gap of the order parameter amplitude mode. It is useful to examine the spectrum of fluctuations of our $r$-fields around the vacuum solution of equations \eqref{eq:stochastic_sys_v2} in the mean field limit setting the sources $s_r$ and $s_a$ for the order parameter to zero. To proceed, we consider perturbative fluctuations around the background with $B_{r\,0}=B_{r\,i}=0$ and $\hat{\psi}_r=\psi_0$. The analysis is very similar to that of \cite{Donos:2022qao} revealing a gapped mode along with a pair of propagating sound modes for wavevectors much smaller than the gap,
\begin{align}\label{eq:spectrum}
    \omega_H&=-i\,\omega_g-i\,D_H\,k^2+\cdots\,,\nn \omega_{\pm}&=\pm\,\sqrt{\frac{\chi_{JJ}}{\chi_b}}\,k-i\,D_s\,k^2+\cdots\,,
\end{align}
where we have set,
\begin{align}\label{eq:omegag}
\omega_g&=-2\,\frac{c_6^R}{|c_6|^2}\,|\psi_0|^2\,u_0\,\frac{\chi_b}{\chi_n}=\frac{c_6^R}{|c_6|^2}\,\frac{4\,\Delta E_0}{|\psi_0|^2}\,,\nn
\chi_b&=\chi_n+\frac{\kappa_0^2}{u_0}=\chi_n\,\frac{\Delta E_0}{\Delta F_0}\,.
\end{align}
In the above equations, $\chi_b$ is the charge susceptibility of the broken phase \cite{Donos:2022qao}, $\Delta E_0$ and $\Delta F_0$ are the mean field energy and the free energy difference between the broken and normal phase. The low energy limit we wish to derive, by integrating out the gapped, amplitude mode $\omega_H$, captures the sound mode $\omega_\pm$. We will postpone an explicit expression for $D_s$ until we have written down the low energy limit theory.

We introduce a perturbative parameter $\lambda$ and take the time and spatial derivatives to be of order $\partial_t\approx\partial_i\approx\mathcal{O}(\lambda)$\footnote{ We can think of the $\lambda$ expansion as being performed on top of the $\varepsilon$ expansion in \eqref{eq:epsilon_exp}. With the initial $\varepsilon$ expansion we construct a theory valid for energies up to the gap of the amplitude mode, which is set by $\varepsilon$. With the introduction of $\lambda$ essentially we focus on energies much smaller than the gap.}.  The mean field theory part of the stochastic system of equations \eqref{eq:stochastic_sys_v2} suggests that our $r$-fields scale according to,
\begin{align}\label{eq:lambda_1}
\hat{\psi}_r&=\rho_r+\delta\rho_r+i\,q\,\rho_r\,\delta\theta_r+\cdots\,,\nn
\delta\rho_r&\approx \delta\theta_r\approx B_{r\,\mu}\approx\mathcal{O}(\lambda)\,,
\end{align}
where we have taken the background VEV $\psi_0=\rho_r$ of the complex scalar to be real.

As we show in Appendix \ref{app:amplitude}, by expanding the generating functional \eqref{eq:generating_eft} to order $\mathcal{O}(\lambda^4)$, we can perform the path integration over the fields $\delta\rho_r$, $\delta\theta_r$ and $\hat{\psi}_a$ to find the low energy effective Lagrangian density,
\begin{align}\label{eq:sf_effective_action}
&\mathcal{L}_{sf}=\rho_b\,C_{a\,0}-c_5\left(\frac{i}{\beta}\,C_{a\,i}^2-C_{a\,i}\,\partial_t C_{r\,i}\right)-\chi_{JJ}\,C_{r}{}^i\,C_{a\,i}\nn
&\quad+\chi_b\,C_{a\,0}\,C_{r\,0}+\chi_b^2\,\zeta_3\,\left(\frac{i}{\beta}\,C^2_{a\,0}-C_{a\,0}\,\partial_t C_{r\,0}\right)\,,
\end{align}
with the gauge invariant vectors  $C_{r,a}=\partial\varphi_{r,a}+A_{r,a}$. We have used the identification \eqref{eq:omegag} and defined the broken phase charge density and current susceptibility,
\begin{align}\label{eq:rho_chi}
    \rho_b=\rho_n-\kappa_0\,\rho_r^2\,,\quad \chi_{JJ}=2\,w_0\,q^2\,\rho_r^2\,.
\end{align} 
Following the details in Appendix \ref{app:amplitude}, the angle $\varphi_r$ coincides with the phase of the original order parameter variable $\psi_r$ that we introduced in section \ref{sec:formalism}. The real constant $\zeta_3$ can be expressed as,
\begin{align}\label{eq:zeta3}
    \zeta_3&=\frac{(\chi_b\,u_0+q\,\kappa_0\,c_6^I)^2+q^2\,\kappa_0^2\,(c_6^R)^2}{\omega_g\,q^2\,\chi_b\,\chi_n\,|c_6|^2\,u_0}\,.
\end{align}

Our low energy degrees of freedom follow the same gauge transformation rules with the phase variables $\phi_r$ and $\phi_a$ in equation \eqref{eq:trans_col}. However, the resulting action does not possess an analog of the time independent gauge symmetry \eqref{eq:gtrans_t_indep}. It is interesting to point out that the hydrodynamic frame we seem to have landed in is the same with the natural frame of holography \cite{Donos:2021pkk}.

Following the details in Appendix \ref{app:amplitude} we can show that the dispersion relations of the perturbative modes of our low energy theory yield the mode $\omega_{\pm}$ of equation \eqref{eq:spectrum} with,
\begin{align}\label{eq:diffusion_constant}
    D_s=\frac{1}{2\,\chi_b}\left(\chi_{JJ}\,\chi_b\,\zeta_3-\,c_5 \right)\,.
\end{align}
According to the $\varepsilon$ expansion of section \ref{sec:formalism}, the bulk viscosity $\zeta_3$ in equation \eqref{eq:zeta3} is of order $\mathcal{O}(\varepsilon^{-2})$. This is certainly to be expected since we have integrated out a mode whose gap behaves like $\mathcal{O}(\varepsilon^{2})$ close to the phase transition. However, the constant $D_s$ remains finite since the current-current susceptibility behaves like $\mathcal{O}(\varepsilon^{2})$ close to the transition as we can see from the expression  \eqref{eq:rho_chi}.

In the normal phase, the original effective theory has two gapped modes \cite{Donos:2022qao} as well as a gapless mode describing charge diffusion with diffusion constant $D_e=-c_5/\chi_n$. This suggests a rearrangement of the degrees of freedom across the phase transition, in agreement with earlier observations concerning holographic models \cite{Donos:2021pkk,Donos:2022qao}. An interesting observation is that the product $\chi_{JJ}\zeta_3$ in equation \eqref{eq:diffusion_constant} remains finite close to the transition. Given that statistical fluctuations become strong near the critical point, it would be interesting to explore how non-linearities could affect this conclusion.\\

\section{Discussion}\label{sec:discussion3}

We constructed the effective theory for the critical dynamics close to a superfluid phase transition in the framework of the Keldysh-Schwinger formalism. An important ingredient to match with Model F \cite{RevModPhys.49.435} as well as with holography \cite{Donos:2022qao} was the ``chemical shift'' symmetry \cite{Dubovsky:2011sj,Crossley:2015evo} of equation \eqref{eq:gtrans_t_indep}, which excluded a number of terms from the effective action and which would otherwise be  allowed in our $\varepsilon$ expansion of the free energy in equation \eqref{eq:fe_expansion}.

In this paper we focused entirely on the coupled sector of the complex order parameter and the corresponding conserved current. This has simplified the problem as it allowed us to focus on  some of the crucial aspects of the construction. This is in direct analogy with the probe limit of holographic theories which decouple the metric fluctuations from the charged and condensed degrees of freedom. However, at finite density this sector also couples to the energy-momentum of the system leading to a much larger description involving the local temperature and normal fluid velocity. We will leave this construction for  future investigation.

As we observed at the end of section \ref{sec:superfluidity}, even though the superfluid effective theory is expected to break down close to the phase transition, the dispersion relations \eqref{eq:spectrum} remain finite. These conclusions were drawn from the superfluid phase point of view leaving open the question of the effect of higher order noise interactions \cite{Chen-Lin:2018kfl,Jain:2020zhu}.

\section*{Acknowledgements}

We would like to thank M. Baggioli for discussions and collaboration on similar topics. We would also like to especially thank N. Iqbal for useful comments on the draft. AD and PK are supported by the Leverhulme Research Project Grant RPG-2023-058. AD is supported by STFC grant ST/T000708/1.

\appendix
\section{Effective Action Construction}\label{app:eff_action_con}


In this Appendix we will discuss some of the details for the derivation of the first couple of terms in the effective Lagrangian expansion \eqref{eq:L_EFT_exp}. We start by discussing the source $s_i$ of the complex scalar $\hat{\psi}_i$ and how it appears in equation \eqref{eq:L_1}. The reason we would like to include such sources in our description is twofold. The first and most obvious is that they would allow us to compute correlation functions for the order parameter. The second is that static, background sources will allow us to study the effects of explicit symmetry breaking in our system. In order to preserve the gauge symmetry transformations, the source $s$ needs to transform in the same way with the order parameter $\psi_i$ according to,
\begin{align}\label{eq:s_gauge_trans}
    s^\prime_i=e^{iq\lambda_i}\,s_i\,,\qquad s^{\star\,\prime}_i=e^{-iq\lambda_i}\,s^\star_i\,.
\end{align}
It is also useful to note that under a KMS transformation, the complex scalar sources transform as,
\begin{align}\label{eq:eq:kms_higgs_s}
    \tilde{s}_r(-x)=s^\star_r(x)\,,\qquad \tilde{s}_a(-x)=s^\star_a(x)+i\beta\,\partial_t s^\star_r(x)\,.
\end{align}

For the purposes of our paper we will only be interested in perturbations of the complex scalar sources and we will only include terms which are linear in $s_r$ and $s_a$. The first property listed in equation \eqref{eq:I_eft_properties_ar} suggests that we can write,
\begin{align}\label{eq:L_1}
\mathcal{L}^{[1]}=&a^\mu\,B_{a\,\mu}+2\,\mathrm{Re}\left[a_\psi\,\hat{\psi}_a+e^{iq\phi_r}s_r^\star\,\hat{\psi}_a\right.\nn
&\quad \left.+e^{-iq\phi_r}\,\hat{\psi}_r^\star s_a+iq\,e^{iq\phi_r}\,\hat{\psi}_r\,s_r^\star\,\phi_a\right]\,,
\end{align}
with the coefficients $a_\psi$ and $a^\mu$ being functions of the $r$-fields. This form satisfies all properties of equation \eqref{eq:I_eft_properties_ar}, provided that $a^\mu$ is real. Notice that the terms involving the complex scalar source $s_{r,a}$ are invariant under the symmetry transformations \eqref{eq:trans_col},\eqref{eq:gtrans_t_indep_newvars} and \eqref{eq:s_gauge_trans} as well as the KMS transformations \eqref{eq:kms_higgs} and \eqref{eq:eq:kms_higgs_s}. Imposing that the rest of the expression \eqref{eq:L_1} is invariant under the KMS transformations \eqref{eq:kms_higgs} gives that,
\begin{align}
a_\psi=\frac{\delta S_0}{\delta \hat{\psi}_r}\,, \quad a^\mu=\frac{\delta S_0}{\delta B_{r\, \mu}}\,,
\end{align}
for some functional $S_0$ of our $r$-fields,
\begin{align}
    S_0[\hat{\psi}_r,\hat{\psi}_r^\star,\phi_r;A_{r\,\mu}]=- \int d^{d-1} x\, F\,.
\end{align}
The final step is to ensure that \eqref{eq:L_1} is also invariant under the transformations given in equations \eqref{eq:gtrans_t_indep_newvars} and \eqref{eq:s_gauge_trans}. Considering first derivatives of our fields, it is easy to check that any function $F$ of the invariant quantities\footnote{In fact, we could include gauge invariant scalars of the form $F_{r\,ij}F_r^{ij}$ with $F_{r\,ij}=\partial_i B_{r\,j}-\partial_j B_{r\,i}$. However, this scalar depends entirely on the source gauge field and we ignore it since we don't consider external magnetic fields.},
\begin{align}\label{eq:symmetry_invariants}
    \hat{\psi}_r\,\hat{\psi}^\star_r\,,\quad D_i\hat{\psi}_r \,D^i\hat{\psi}^\ast_r\,,\quad B_{r\,0} \,,
\end{align}
satisfy our criteria.  This allows us to write the most general function that satisfies our constraints as,
\begin{align}\label{eq:F_T}
    F=F\left(\hat{\psi}_r\,\hat{\psi}^\star_r, D_i\hat{\psi}_r \,D^i\hat{\psi}^\ast_r ,B_{r\,0}\right)\,.
\end{align}

Before constraining the function $F$ further, we will turn our attention to the second term in the expansion of the effective action in \eqref{eq:L_EFT_exp}. The most general expression with single time derivative terms we can write and which are invariant under the transformations \eqref{eq:gtrans_t_indep_newvars} reads,
\begin{align}
\mathcal{L}^{[2]}=&i\, c_1 \,\hat{\psi}_a\hat{\psi}_a^\star+i\, c_2\, B_{a\, 0}^2+i c_3\, B_{a\, i}^2+c_4\, B_{a\, 0}\,\partial_t B_{r\,0}\notag\\
&+c_5\, B_{a\, i}\, \partial_t B_{r\, i}+c_6\, \hat{\psi}_a\, \partial_t \hat{\psi}_r^\star+c_6^\star\, \hat{\psi}_a^\star\, \partial_t \hat{\psi}_r\,.
\end{align}
The first line of the properties in equation \eqref{eq:I_eft_properties_ar} demand that $c_1,\ldots,c_5$ are real and $c_6=c_6^R+i\,c_6^I$ can be complex. Moreover, these can be functions of the invariant quantities \eqref{eq:symmetry_invariants}. Imposing KMS invariance, one can easily show that these coefficients are constrained in a way such that,
\begin{align}\label{eq:L_2}
&\mathcal{L}^{[2]}=-\frac{2\,i\, c_6^R}{\beta} \hat{\psi}_a\hat{\psi}_a^\star-i \frac{c_4}{\beta} B_{a\, 0}^2+c_4\, B_{a\, 0}\,\partial_t B_{r\, 0}\notag\\
&\,\,-i \frac{c_5}{\beta} B_{a\, i}^2+c_5\, B_{a\, i}\, \partial_t B_{r\, i}+2\,\mathrm{Re}\left(c_6\, \hat{\psi}_a\, \partial_t \hat{\psi}_r^\star\right)\,.
\end{align}
Finally, imposing the third property of equation \eqref{eq:I_eft_properties_ar}, we conclude that our dissipative coefficients must satisfy the inequalities $c_6^R, c_4, c_5 \leq 0$.

The off-shell conserved current reads,
\begin{align}\label{eq:currents_offshell_2}
    \hat{J}_r^i&=\frac{\delta I_{EFT}}{\delta A_{a\,i}}=J_r^i-i\frac{2\,c_5}{\beta}B_{a}{}^i\notag\\
    &=-2q\,\mathrm{Im}\left(\hat{\psi}^\star_r\,\frac{\partial F}{\partial D_i\hat{\psi}^\star_r}\right)+c_5\,\partial_t B_{r}{}^i-i\frac{2\,c_5}{\beta}B_{a}{}^i\,,\notag\\
    \hat{J}_r^0&=\frac{\delta I_{EFT}}{\delta A_{a\,0}}=J_r^0-i\frac{2\,c_4}{\beta} B_{a\,0}\notag\\
    &=-\frac{\partial F}{\partial B_{r\,0}}+c_4\,\partial_t B_{r\,0}-i\frac{2\,c_4}{\beta} B_{a\,0}\,,
\end{align}
where $J_r^\mu$ denotes the classical part of the electric current. The above shows that in thermodynamic equilibrium the charge density is given by the derivative of the function $-F$ with respect to $B_{r\,0}$. In order to understand the role of the function $F$, it will be enlightening to consider the classical equations of motion of our system. These can obtained from the effective action by taking derivatives with respect to the $a$-fields and setting them equal to zero. Doing so reveals the the derivatives of $F$ with respect to the complex order parameter is fixed by the classical source $s_r$. The above show that we can treat $F$ as the Ginzburg-Landau-Wilson potential since $\mu=B_{r\,0}$ is the chemical potential of the system. This suggests that, at the level of thermodynamics, the energy density of the system is $E=F+\mu\,J_r^0$.

Expanding the function $F$ in powers of the order parameter near criticality we obtain,
\begin{align}\label{eq:F_gen_form}
    F=&F_0(\mu)+r(\mu)\,|\hat{\psi}_r|^2+\frac{1}{2}u(\mu)\,|\hat{\psi}_r|^4 \nn
    &\quad +w(\mu)\,D_i\hat{\psi}_r\,D^i\hat{\psi}^\star_r+\cdots\,.
\end{align}
In the above expansion we introduced the constant $F_0$, which is identified as the normal phase free energy. Its first derivative gives minus the normal phase charge density $\rho_n$ and its second derivative will therefore yield minus the susceptibility of the normal phase $\chi_n$, so that in a semi-classical approximation,
\begin{align}
    \rho_n=-\left.\frac{\partial F_0}{\partial \mu}\right|_{\mu=\mu_0}\,,\qquad \chi_n=-\left.\frac{\partial^2 F_0}{\partial \mu^2}\right|_{\mu=\mu_0}\,,
\end{align}
where $\mu_0$ is the value of the chemical potential in the thermal state.

It is useful to note that our functions $F_0$, $r$, $u$ and $w$ will in general depend on temperature as well as the the deformation parameters and coupling constants of the microscopic theory. For example, one can imagine that the system is deformed by a relevant operator which doesn't have to be included in our low energy description. However, the corresponding deformation parameters will in general enter in the effective action.

Being interested in the dynamics of our system close to the phase transition, it is reasonable to perform the shift,
\begin{align}
    B_{r\,0}\rightarrow \mu_0+B_{r\,0}\,,
\end{align}
allowing us to treat $B_{r\,\mu}$ as a fluctuation around the thermal state. At the same time, close to the transition we will take our  derivatives, fields and constants to scale according to equation \eqref{eq:epsilon_exp}. The KMS transformation rules \eqref{eq:kms_higgs} then give the $a$-field scaling rules,
\begin{align}
    \phi_{a}\propto\mathcal{O}(\varepsilon^2)\,,\quad \hat{\psi}_a\propto \mathcal{O}(\varepsilon^3)\,, \quad s_a\propto \mathcal{O}(\varepsilon^5)\,.
\end{align}
Assuming that the constant $c_4$ behaves as a regular function of $\varepsilon$, the above scalings suggest that we can drop the corresponding term in \eqref{eq:L_2} by setting $c_4=0$. This leads us to the expression quoted in equation \eqref{eq:L_2_exp}.

Keeping terms up to order $\varepsilon^4$ in $F$ (or order $\varepsilon^6$ in $I_{EFT}$) we arrive at the expression of equation \eqref{eq:fe_expansion}. It is useful to understand the bare constants that appear in the expansion of equation \eqref{eq:fe_expansion} for the function $F$ in the context of mean field theory. Demanding that the free energy is extremised by the mean field value $\psi_0$ of the order parameter in the undeformed theory, we can identify,
\begin{align}\label{eq:u_0}
    r_0=\frac{2\,\Delta F_0}{|\psi_0|^2}\,,\qquad u_0=-\frac{2\,\Delta F_0}{|\psi_0|^4}\,,
\end{align}
where $\Delta F_0$ is the free energy density difference between the broken and the normal phase. In terms of mean field theory, our previous arguments lead to,
\begin{align}\label{eq:kappa}
    \kappa_0=\left.\frac{\partial r}{\partial\mu}\right|_{\mu=\mu_0}=-\frac{\Delta\rho_0}{|\psi_0|^2}\,.
\end{align}
In the above expression, $\psi_0$ is the mean field value of the order parameter in the broken phase and $\Delta\rho_0$ is the charge density difference between the broken and the normal phase close to the transition.

For convenience, it is useful to note that the energy density difference between the broken and the normal phase is,
\begin{align}\label{eq:energy_difference}
    \Delta E_0=\Delta F_0-\frac{1}{2\,\chi_n}\,(\Delta\rho_0)^2\,,
\end{align}
where on the left hand side we have fixed charge density and on the right hand side we have fixed background chemical potential $\mu_0$.

\section{Derivation of the Stochastic System}\label{app:stochastic}


In this Appendix we will give some of the details needed to derive the stochastic equations of motion that we have quoted in the main text in equation \eqref{eq:stochastic_sys_v2} along with the correlation functions for the noise fields.

In order to find the equations of motion with the appropriate noise terms from the effective theory of section \ref{sec:formalism}, we will make use of  the following well-known identity \footnote{ This is simply a path integral generalisation of the one-dimensional Gaussian integral.},

\begin{align}\label{eq:pi}
    &\int D\phi\, e^{-\int d^dx\,d^dy\,\phi(x)\,K(x,y)\,\phi(y)+i\int d^dx\, J(x)\,\phi(x)}=\nn& \quad\det\left(\frac{K}{\pi}\right)^{-\frac{1}{2}}\,e^{-\frac{1}{4}\int d^dx\,d^dy\,J(x)\,K^{-1}\,(x,y)\,J(y)}\,.
\end{align}
In what follows, numerical constants such as the determinant factor in the above expression will be absorbed in the integration measure as they carry no dependence on the interesting part which is the sources $J(x)$.

In order to obtain the stochastic equation of motion \eqref{eq:stochastic_sys_v2} for the complex order parameter $\hat{\psi}_r$, we first split the field $\hat{\psi}_a$ in real and imaginary parts: $\hat{\psi}_a=\hat{\psi}^R_a+i\,\hat{\psi}^I_a$. We then introduce two real fields $z_1,z_2$ and apply\eqref{eq:pi} to obtain,
\begin{align}
e^{\int d^dx\, \frac{2\,c_6^R}{\beta}(\hat{\psi}^R_a)^2}&=\int Dz_1\, e^{\int d^dx\, \frac{\beta}{2\, c_6^R }z_1^2+2\,i\,\hat{\psi}^R_a\, z_1}\,,\nn
e^{\int d^dx\, \frac{2\,c_6^R}{\beta}(\hat{\psi}^I_a)^2}&=\int Dz_2\, e^{\int d^dx \frac{\beta}{2\,c_6^R }z_2^2+2\,i\,\hat{\psi}^I_a z_2}\,.
\end{align}
Using these identities, $\hat{\psi}_a$ appears linearly in the effective action. As a result, the path integral over $\hat{\psi}_a$ and $\hat{\psi}_a^\star$ gives two delta functions, which constrain the r-fields $\hat{\psi}_r,\hat{\psi}_r^\star$ to be on-shell, obeying the complex scalar stochastic equation,

\begin{align}
 c_6^\star\, \partial_t \hat{\psi}_r&=-\frac{\delta S_0}{\delta \hat{\psi}_r^\star}-e^{-iq\phi_r}\,s_r-z\,.
\end{align}

The complex noise field $z=z_1+i\,z_2$ then satisfies,

\begin{align}
\langle z(x)\,z(y)\rangle &=\int Dz\,Dz^\star z(x)\,z(y)\,e^{\int d^dx\, \frac{\beta}{2\, c_6^R}|z|^2}=0\,,\nn
  \langle z^\star(x)\,z(y)\rangle &=\int Dz\,Dz^\star z^\star(x)\,z(y)\,e^{\int d^dx\, \frac{\beta}{2\, c_6^R}|z|^2}\nn&=-\frac{2\,\mathrm{Re}(c_6)}{\beta}\,\delta^{(d)}(x-y)\,.
\end{align}

 For the noise related to the $\phi_a$ field we have essentially two choices. The first one is to introduce a real vector noise field $\zeta^\mu$ according to the identities,
\begin{align}
    e^{\int d^dx \frac{c_4}{\beta}B_{a\,0}^2}=\int D\zeta^0 \, e^{\int d^dx\,\frac{\beta}{4\,c_4}\zeta_0^2+i\,\zeta^0 B_{a\,0}}\,,\nn
      e^{\int d^dx \frac{c_5}{\beta}B_{a\,i}^2}=\int D\zeta^i \, e^{\int d^dx\,\frac{\beta}{4\,c_5}\zeta_i^2+i\,\zeta^i B_{a\,i}}\,.
\end{align}
The variable $\phi_a$ then appears linearly in the effective action and can be integrated over, yielding a delta functional with argument,
\begin{align}\label{eq:delta_curr}
&-\partial_\mu\,J_r^\mu-\partial_\mu\zeta^\mu+2\,\mathrm{Re}(i\,q\,e^{iq\phi_r}\hat{\psi}_rs_r^\star)\,,
\end{align}
with $J_r^\mu$ as defined in equation \eqref{eq:currents_offshell_2}. However, the weight of the path integral now involves the ``source'' term factor $e^{\int d^dx\,i\,\zeta^\mu\,A_{a\,\mu}}$. To absorb it, we can make the shift,
\begin{align}
\zeta^0\to\zeta^0-\frac{2\,i\,c_4}{\beta}\,A_{a\,0}\,,\nn
\zeta_i\to\zeta_i-\frac{2\,i\,c_5}{\beta}\,A_{a\,i}\,.
\end{align}
The argument of the delta functional \eqref{eq:delta_curr} implies then the current continuity equation with noise,

\begin{align}
      \partial_\mu J^\mu_r&=-2\,q\,\mathrm{Im}\left[ \hat{\psi}_r\,e^{iq\phi_r}\,s_r^\star\right] -\partial_\mu\zeta^\mu\nn & \qquad+\frac{2\,i}{\beta}\left(c_5\,\partial_i A_a^i+c_4\,\partial_t A_{a\,0} \right)\,.
\end{align}

For the correlation function of the noise field we have,
\begin{align*}
    \langle \zeta^0(x)\,\zeta^0(y) \rangle&=\int D\zeta^\mu\, \zeta^0(x)\,\zeta^0(y)\, e^{\int d^dx\,\frac{\beta}{4\,c_4}(\zeta^0)^2}\nn&=-\frac{2\,c_4}{\beta}\delta^{(d)}(x-y)\,,\nn
      \langle \zeta^i(x)\,\zeta^j(y) \rangle&=\int D\zeta^\mu\, \zeta^i(x)\,\zeta^j(y)\, e^{\int d^dx\,\frac{\beta}{4\,c_5}\zeta_k^2}\nn&=-\delta^{ij}\,\frac{2\,c_5}{\beta}\delta^{(d)}(x-y)\,.
\end{align*}

An alternative way to write the stochastic equation for the current is to introduce a scalar noise field $\zeta$ through the identity,
\begin{align}
    &e^{\int d^dx\, \phi_a (-\frac{c_4}{\beta}\partial_t^2-\frac{c_5}{\beta}\partial_i^2)\phi_a}=\nn &\int D\zeta\,e^{\int d^dx\,( -\frac{\beta}{4}\,\zeta\,(c_4\,\partial_t^2+c_5\,\partial_i^2)^{-1}\zeta+i\,\zeta\,\phi_a)}\,.
\end{align}
The integral over $\phi_a$ constraints the current to be on-shell, obeying the constraint equation of motion,
\begin{align}
     \partial_\mu J^\mu_r=&-2\,q\,\mathrm{Im}\left[ \hat{\psi}_r\,e^{iq\phi_r}\,s_r^\star\right] +\zeta\nn&\quad\quad+\frac{2\,i}{\beta}\left(c_5\,\partial_i A_a^i+c_4\,\partial_t A_{a\,0} \right)\,,
\end{align}
with the correlation function for the scalar noise field obeying,
\begin{align}
\langle \zeta(x)\zeta(y)\rangle =\frac{2}{\beta}(c_4\,\partial_t^2+c_5\,\partial_i^2)\delta^{(d)}(x-y)\,,
\end{align}
and $J^\mu_r$ as defined in equation \eqref{eq:currents_offshell_2}.\\

\section{Integrating out the Amplitude Mode}\label{app:amplitude}


In this Appendix we provide some of the necessary technical details regarding the derivation of section \ref{sec:superfluidity}.

For convenience, we introduce the real and imaginary parts,
\begin{align}
    \hat{\psi}_a=\hat{\psi}_a^R+i\,\hat{\psi}_a^I\,,
\end{align}
obeying the perturbative KMS transformation rule,
\begin{align}\label{eq:kms_higgsv2}
    \tilde{\hat{\psi}}_a^R(-x)&=\hat{\psi}_a^R+i\,\beta\,\partial_t\delta\rho_r\,,\nn
    \tilde{\hat{\psi}}_a^I(-x)&=-\hat{\psi}_a^I-i\,\beta\,q\,\rho_r\,\partial_t\delta\theta_r\,.
\end{align}

Consistency of the KMS transformations \eqref{eq:kms_higgsv2} then leads to,
\begin{align}\label{eq:lambda_2}
    \hat{\psi}_a\approx \hat{\psi}_a^\star\approx B_{a\,\mu}\approx \mathcal{O}(\lambda^2)\,.
\end{align}
Based on the discussion around equations \eqref{eq:lambda_1} and \eqref{eq:lambda_2}, we can expand the effective action in the parameter $\lambda$. Indeed, the derivative expansion scheme above has similar flavour to a semi-classical approximation. In this expansion we aim to retain quadratic terms in $a$-fields, suggesting that we maintain terms up to order $\mathcal{O}(\lambda^4)$, to obtain,
\begin{align}
\mathcal{L}^{[1]}=&\left(-4\,u_0\,\rho_r^2\,\delta\rho_r-2\,\kappa_0\,\rho_r\,B_{r\,0}\right)\,\hat{\psi}^R_a+\left(\rho_n-\kappa_0\,\rho_r^2-2\,\kappa_0\,\rho_r\,\delta\rho_r+\chi_n\,B_{r\,0} \right)\,B_{a\,0}\nn +&2\,w_0\,q\,\rho_r\, \partial^i\left(\partial_i\delta\theta_r+B_{r\,i}\right)\,\hat{\psi}^I_a
-2\,w_0\,q^2\,\rho_r^2\,\left(\partial^i\delta\theta_r+B_r{}^i\right)\,B_{a\,i}\label{eq:L1_exp}\,,\\
\mathcal{L}^{[2]}=&2\,\left(c_6^R\,\hat{\psi}_a^R-c_6^I\,\hat{\psi}_a^I \right)\partial_t\delta\rho_r +2\,q\,\rho_r  \left(c_6^R\,\hat{\psi}_a^I+c_6^I \,\hat{\psi}_a^R \right)\partial_t\delta\theta_r-\frac{2\,i\,c_6^R}{\beta}\left[ \left(\hat{\psi}_a^R\right)^2+\left(\hat{\psi}_a^I\right)^2\right]\nn
-&\frac{i\,c_5}{\beta}\,B_a{}^i\,B_{a\,i}+c_5\,B_a{}^i\,\partial_t B_{r\,i}\,.\label{eq:L2_exp}
\end{align}


We now observe that $\delta\rho_r$ appears linearly in the effective action terms \eqref{eq:L1_exp} and \eqref{eq:L2_exp}. Performing the path integration over this variable yields a delta functional imposing a constraint which can be solved perturbatively up to third order in $\lambda$ according to,
\begin{align*}
    \hat{\psi}^R_a=-\frac{\kappa_0}{2\,u_0\,\rho_r}B_{a\,0}+\frac{\kappa_0\,c_6^R}{4\,u_0^2\,\rho_r^3}\partial_t B_{a\,0}+\frac{c_6^I}{2\,u_0\,\rho_r^2}\partial_t \hat{\psi}^I_a\,.
\end{align*}
In order for this to make sense, we must ensure that the domain of integration in the path integral does not include the kernel of the operator we are trying to invert. This is guaranteed by the fact that we are assuming that the gap of the mode with dispersion relation $\omega_H$ in equation \eqref{eq:spectrum} is larger than the UV cut off scale $\Lambda$ of our effective theory. By substituting the above in the effective action we find the expressions,
\begin{align}
    \mathcal{L}^{[1]}=&B_{r\,0}\Big(\chi_b\,B_{a\,0}-\frac{\kappa_0\,c_6^I}{u_0\,\rho_r}\partial_t\hat{\psi}^I_a-\frac{\kappa_0^2\,c_6^R}{2\,u_0^2\,\rho_r^2}\partial_t B_{a\,0}\Big)+\rho_b\, B_{a\,0}-2\,w_0\,q^2\,\rho_r^2(\partial_i \delta\theta_r+B_{r i})B_{a\, i}\nn
    &-2\,w_0\,q\,\rho_r (\partial_i \delta\theta_r+B_{r\,i})\partial_i\hat{\psi}^I_a\,,\\
    \mathcal{L}^{[2]}=&-\frac{2\,i\,c_6^R}{\beta}(\hat{\psi}_a^I)^2-i\frac{c_6^R \,\kappa_0^2}{2\,\beta\,u_0^2\,\rho_r^2} B_{a\,0}^2+2\,q\,\rho_r \,c_6^R\,\hat{\psi}_a^I \partial_t\delta\theta_r-\frac{\kappa_0 \, c_6^I\,q}{u_0}B_{a\,0}\,\partial_t \delta\theta_r\nn-&\frac{i\,c_5}{\beta}\,B_a{}^i\,B_{a\,i}+c_5\,B_a{}^i\,\partial_t B_{r\,i}\,.
\end{align}
In the above expressions we have defined $\chi_b$ and $\rho_b$, the charge susceptibility and the charge density respectively of the broken phase, given by,
\begin{align}
    \chi_b=\chi_n+\frac{\kappa_0^2}{u_0}\,,\qquad
    \rho_b=\rho_n-\kappa_0\,\rho_r^2\,.
\end{align}
At this point, it is useful to make a change of variables  in the path integral \footnote{These changes are just shifts of the integrated variables and so have unit Jacobian.}:
\begin{align}
    \phi_r \to \varphi_r&=\phi_r+\delta\theta_r\,,\nn
    \phi_a \to \varphi_a&=\phi_a+\frac{1}{q\,\rho_r}\hat{\psi}^I_a\,.
\end{align}
As in the main text, we introduce the gauge invariant vectors  $C_{r}=\partial\varphi_r+A_r$ and $C_a=\partial\varphi_a+A_a$. Note that the last  term of $\mathcal{L}^{[1]}$ will change by a total time derivative term, which we will drop.

The next step is to observe that the variable $\delta \theta_r$ appears linearly in the resulting effective action as well. Performing the corresponding path integration over it we obtain the delta functional,
\begin{align*}
   &\delta \left(-2\,q\,\rho_r\,c_6^R\,\partial_t \hat{\psi}^I_a +(\chi_b+\frac{q\,\kappa_0}{u_0}c_6^I)\partial_t C_{a\,0}\right)=\nn
   & \frac{1}{\mathrm{det}(-2\,q\,\rho_r\,c_6^R\,\partial_t)}\delta \Big(\hat{\psi}^I_a-\frac{1}{2\,q\,\rho_r c_6^R}(\chi_b+\frac{q\,\kappa_0}{u_0}c_6^I) C_{a\,0}\Big)\,.
\end{align*}
For the equality to make sense, $\partial_t$ acting on $\hat{\psi}^I_a$ has to be an invertible operator. The reason this is the case, is because its domain of definition contains only functions that vanish at $t \to +\infty$. This is true for all $a$ fields due to the boundary condition \eqref{eq:bc_infinity}. As a result, the $\partial_t$ has an empty kernel on the space of functions we are integrating over. The above shows that the operation is meaningful. The final step is to integrate over $\hat{\psi}^I_a$ using the delta functional in order to arrive to our final result of equation \eqref{eq:sf_effective_action} for the effective action, after identifying the current susceptibility $\chi_{JJ}$ through \eqref{eq:rho_chi}.

In order to express the constant $\zeta_3$ of equation \eqref{eq:zeta3} in terms of the variables appearing in Model F, we can use the matching relations \eqref{eq:model_f},\eqref{eq:u_0} and \eqref{eq:kappa} to write the expression,
\begin{align}
   \zeta_3= \frac{1}{q^2\,\rho_r^2\,\mathrm{Re}\Gamma_0}\Big( \mathrm{Re}\Gamma_0^2+\Big(\mathrm{Im}\Gamma_0+\frac{q\,\rho_r^2\,\Delta\rho_0}{4\,\chi_b\,\Delta F_0}\Big)^2\Big)\,.
\end{align}
This expression matches precisely the previously obtained result in \cite{Donos:2022qao}. Moreover, we can write the KMS transformation rules for our low energy fields,
\begin{align*}
    \tilde{C}_{r\,\mu}(-x)=C_{r\,\mu}(x)\,,\,\tilde{C}_{a\,\mu}(-x)=C_{a\,\mu}+i\,\beta\,\partial_t C_{r\,\mu}(x)\,,
\end{align*}
and check that the effective action \eqref{eq:sf_effective_action} is indeed invariant.

In order to make contact with classical superfluid hydrodynamics, it is useful to write down the expression for the conserved current,
\begin{align}\label{eq:superlfuid_constitutive}
    \hat{J}_r{}^0=&\frac{\partial\mathcal{L}_{sf}}{\partial C_{a\,0}}=\rho_b+\chi_b\,C_{r\,0}-\chi_b^2\,\zeta_3\,\partial_t C_{r\,0}+\frac{2\,i}{\beta}\,\chi_b^2\,\zeta_3\,C_{a\,0}\,,\nn
    \hat{J}_r{}^i=&\frac{\partial\mathcal{L}_{sf}}{\partial C_{a\,i}}=-\chi_{JJ}\,C_{r}{}^i+c_5\,\partial_t C_r{}^i-\frac{2\,i\,c_5}{\beta}\,C_a{}^i\,,
\end{align}
The mean field theory part of the above constitutive relations is consistent with superfluid hydrodynamics \cite{KHALATNIKOV198270,Herzog:2011ec,Bhattacharya:2011tra,Donos:2021pkk} after identifying $\zeta_3$ with the third bulk viscosity.

\newpage
\bibliographystyle{utphys}
\bibliography{refsthesis}{}
\end{document}